\documentclass[a4paper]{article}

\usepackage{INTERSPEECH2020}
\usepackage{float}
\usepackage{subfig}
\usepackage{booktabs}
\usepackage{url}
\usepackage{hyperref}

\title{Analysis of Emotional Content in Indian Political Speeches}
\name{Sharu Goel$^1$,Sandeep Kumar Pandey$^2$, Hanumant Singh Shekhawat$^3$}
\address{
  $^{1,2,3}$Indian Institute of Technology Guwahati, India}
\email{sharugoel@iitg.ac.in, sandeep.pandey@iitg.ac.in, h.s.shekhawat@iitg.ac.in} 

\begin{document}

\maketitle
\begin{abstract}
Emotions play an essential role in public speaking. The emotional content of speech has the power to influence minds. As such, we present an analysis of the emotional content of politicians speech in the Indian political scenario. We investigate the emotional content present in the speeches of politicians using an Attention based CNN+LSTM network.  Experimental evaluations on a dataset of eight Indian politicians shows how politicians incorporate emotions in their speeches to strike a chord with the masses. An analysis of the voting share received along with victory margin and their relation to emotional content in speech of the politicians is also presented.

\end{abstract}

\section{Introduction}

Speech signal contains information on two levels. On the first level, it contains the message to be conveyed and on the second level, information such as speaker, gender, emotions, etc. Identifying the emotional state of a speech utterance is of importance to researchers to make the Human-Computer Interaction sound more natural. Also, emotions play a significant role when it comes to public speaking, such as a politician addressing a crowd. Speeches offer politicians an opportunity to set the agenda, signal their policy preferences, and, among other things, strike an emotional chord with the public. We can observe that emotional speeches allow the speaker to connect better with the masses, instill faith in them, and attract a strong response. As such, it becomes interesting to analyze which emotions and in what proportion do politicians incorporate in their speeches. 

While voting for a particular politician in elections depends on several factors such as background, work done in the previous tenure, public interaction etc, studies done in \cite{riker1968theory} suggest that emotions or affects have a negative impact on voting decisions of the public. Also, an experiment performed by the University of Massachusetts Amherst revealed that under the emotional state of anger, we are less likely to systematically find information about a candidate and increase our
reliance on stereotypes and other pre-conceived notions and heuristics \cite{parker2010vote}. Another independent
study revealed that anger promotes the propensity to become politically active and hence, has a normatively desirable consequence of an active electorate \cite{weber2013emotions}. Furthermore, a study on the 2008 Presidential Elections in the United States showed that negative emotions like anger and outrage have a substantial effect on mobilizing the electorate, and positive emotions like hope, enthusiasm, and happiness have a weaker mobilizing effect. The researchers excluded emotions like sadness and anxiety from their study, as these emotions did not have a noticeable influence on voting behavior \cite{valentino2011election}. The results of the previous experiments and studies were backed with the outcome of the 2016 Presidential Elections in the United States, where the current president Donald Trump had more anger and hope dominated campaign advertisements compared to his opposition, Hilary Clinton \cite{searles2017use}.

Nevertheless, a question remains; do emotional manipulations elicit the expected emotions? Does an angry speech elicit anger or some other emotion? A study on the structure of emotions\cite{weber2013emotions} in which participants were randomly assigned to view one of four campaign ads designed to elicit a discrete emotion revealed that although there is heterogeneity in emotional reactions felt in response to campaign messages, the emotional manipulations from campaign ads did elicit the emotions expected. The results of the study showed that sadness was most prevalent in the sadness condition relative to all other conditions, anger was higher in the anger condition relative to all other conditions, and enthusiasm was much more significant in the enthusiasm condition relative to all other conditions. Nonetheless, the study also illustrated that different emotional manipulations were effective in eliciting a specific emotion. For instance, the study showed that sadness was felt in response to both angry and sad emotional manipulations. Similarly, anger was felt in response to sad emotional manipulation as well (although anger was more prevalent in angry emotional manipulation, as stated earlier).

Moreover, due to the recent advancements in the deep learning field, SER has seen significant improvement in performance. In \cite{yang2018predicting}, an end-to-end deep convolutional recurrent neural network is used to predict arousal and valence in continuous speech by utilizing two separate sets of 1D CNN layers to extract complementary information. Also, in \cite{li2018attention}, feature maps from both time and frequency convolution sub-networks are concatenated, and a class-based and class-agnostic attention pooling is utilized to generate attentive deep features for SER. Experiments on IEMOCAP \cite{busso2008iemocap} dataset shows improvement over the baseline. Also, researchers have explored the possibility of extracting discriminative features from the raw speech itself. In \cite{wagner2018deep},  a comparison of the effectiveness of traditional features versus end-to-end learning in atypical affect and crying recognition is presented, only to conclude that there is no clear winner. Moreover, works in \cite{pandey2019emotion} and \cite{sarma2018emotion} have also utilized raw speech for emotion classification.

\begin{figure*}[!t]
  \centering
  \includegraphics[ height =6cm , width=\linewidth ]{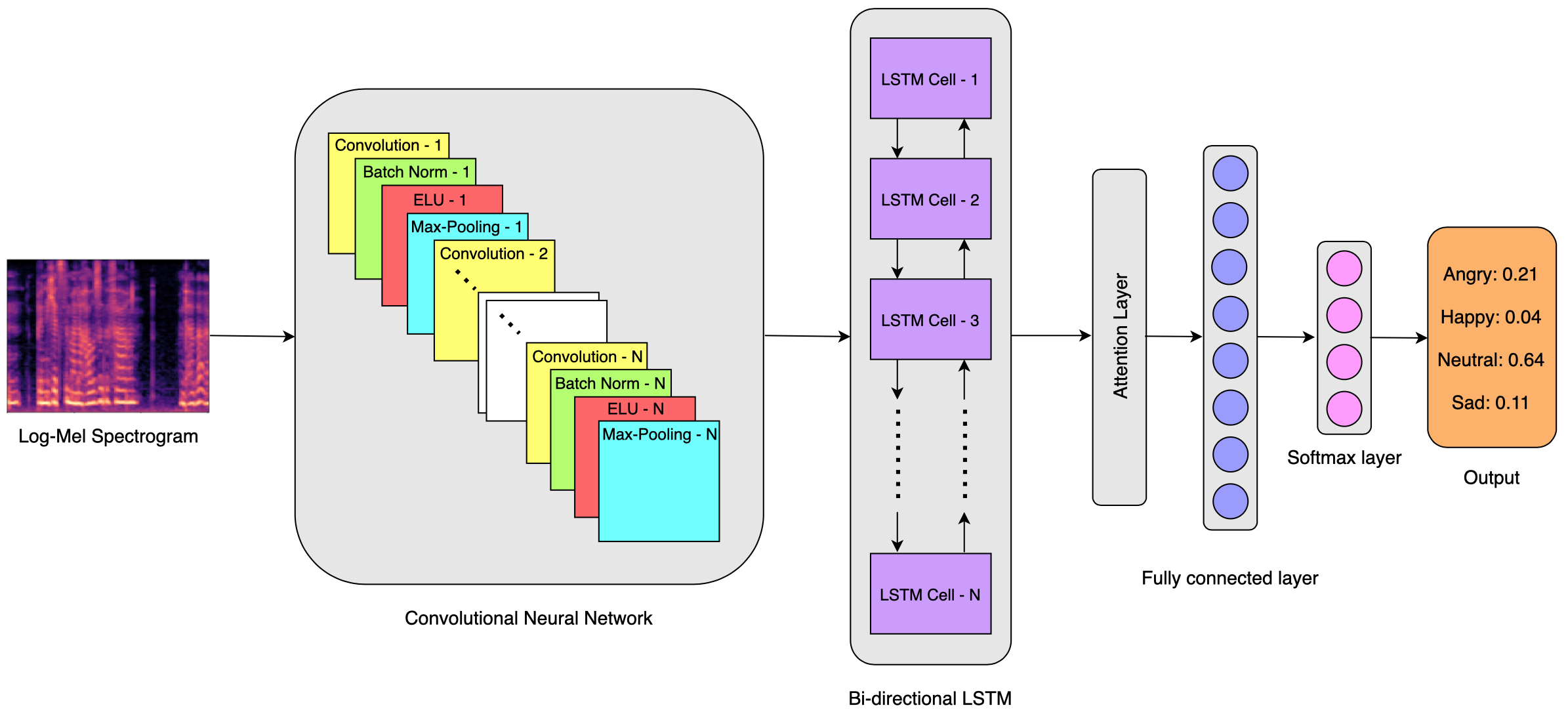}
  \caption{Attentive CNN+LSTM architecture}
  \label{fig:angry_spec}   
\end{figure*}
In this work, we propose a deep learning-based architecture to the task of emotional analysis of politicians speech, particularly in the Indian political context. We present a dataset of speech utterances collected from speeches of 8 Indian politicians. Also, a widely used Attentive CNN+LSTM architecture is used to the task of Speech Emotion Recognition (SER). At the first level, the speech utterances are classified into four emotion categories- Angry, Happy, Neutral, and Sad, using the trained emotion model. At the second level, analysis of the amount of different emotions present in a politician's speech is presented, which gives us a first-hand view of the emotional impact of speeches on the listeners and the voting decisions. To the best of our knowledge, this is the first work in analyzing the emotional contents of speeches of Indian politicians.

The remainder of the paper is organized as follows:  Section 2 describes the data collection strategy and dataset description, along with the perceptual evaluation of the dataset. Section 3 describes the Attentive CNN+LSTM architecture in brief. The experimental setup and results are discussed in Section 4.  Section 5 summarizes the findings and concludes the paper.

\section{Dataset Description}

Since no such speech corpus exists which suits our research in the Indian political context, we present the IITG Politician Speech Corpus. The Politician Speech Corpus consists of 516 audio segments from speeches delivered by Indian Politicians in Hindi. 
The speeches are publicly available on YouTube on the channels of the respective political groups the politicians belong to. Eight Indian politicians, whom we nicknamed for anonymization, namely NI (88 utterances), AH (101 utterances), AL (81 utterances), JG (20 utterances), MH (8 utterances),  RI (31 utterances), RH (122 utterances), SY (65 utterances) were considered as they belonged to diverse ideologies and incorporate diverse speaking style. The politicians were chosen based on the contemporary popularity and wide availability of speeches. The speeches were addressed in both indoor(halls or auditoriums) and outdoor(complex grounds, courts, or stadiums) environment and carried substantial noise in the background. Only audio output from a single channel was considered. 

 The extracted audio from video clips is first segmented into utterances of length 7-15 seconds and classified into four basic emotions - Angry, Happy, Neutral and Sad . Perceptual evaluation of the utterances are performed by four evaluators. For correctly labelling the speech utterances we followed a two step strategy. The first step was to identify if there is any emotional information in the utterance, irrespective of the underlying message. This is done by assigning each utterance a score on a scale of one to five(one meaning not confident and five meaning extremely confident) based on how confidently the evaluators were able to assign it an emotion. The second step was to assign a category label to the utterances . Since the speeches from politicians exhibit natural emotions, it may happen that it is perceived differently by different evaluators. For this we followed the following strategy for assigning a categorical label -
 \begin{enumerate}
     \item If the majority of listeners are in favor of one emotion, then the audio clip was assigned that emotion as its label. The confidence score of the label would be the mean score of the majority.
     \item If two listeners are in favor of one emotion and two in favor of the other, then the emotion with the highest mean confidence score was assigned as the label. The confidence score of the label would be the same mean used to break the tie. 
     \item If no consensus is achieved for a particular utterance i.e. all listeners assigned a different label, then that audio clip is discarded.
 \end{enumerate}{}
 
 The number of utterances in each emotional category along with the confidence score of the evaluators, is presented in Table 1. Happy emotion category has the least number of utterances followed by Sad, which is in line with the literature on the emotional content of politician speeches. This shows that the classes are highly imbalanced, which adds to the difficulty level of the task at hand.

\begin{table}[]
\renewcommand{\arraystretch}{1.3}
\caption{ Distribution of utterances emotion wise along with confidence score for each emotion class generated using perceptual evaluation}
\begin{center}

\resizebox{0.47\textwidth}{!}{
\begin{tabular}{@{}cccc@{}}
\toprule
\textbf{Emotion} & \textbf{No.of Samples} & \textbf{Confidence Score (1-5)} & \textbf{Confidence \%} \\ \midrule
Angry            & 175                    & 3.5519                          & 71.0381                \\
Happy            & 36                     & 3.2152                          & 64.3055                \\
Neutral          & 230                    & 3.8199                          & 76.3991                \\
Sad              & 75                     & 3.7688                          & 75.3778                \\ \bottomrule
\end{tabular}}
\end{center}
\end{table}


 

\section{Architecture Description}

The deep learning architecture used for the classification of emotional states of the utterances is the state-of-the-art CNN+LSTM with attention mechanism \cite{chen20183}, \cite{zhao2019speech}, as shown in Figure 1. The architecture consists of four local feature learning blocks (LFLB), one LSTM layer for capturing the global dependencies, and one attention layer to focus on the emotion relevant parts of the utterance followed by a fully-connected softmax layer to generate class probabilities.

\begin{figure}[!t]
  \centering
  \includegraphics[ height =5cm , width=0.8\linewidth ]{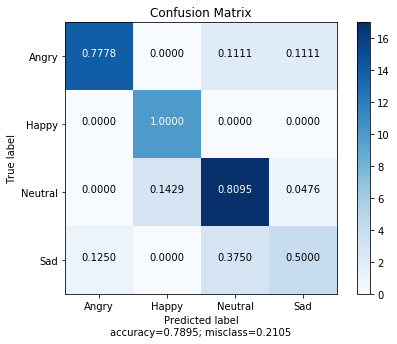}
  \caption{Confusion matrix of Attentive CNN+LSTM architecture, with an average accuracy of $78.95\%$, where the rows represent the confusion of the ground truth emotion during prediction.}
  \label{fig:confusion_mat}   
\end{figure}

Each LFLB comprises of one convolutional layer, one batch-normalization layer, one activation layer, and one max pooling layer. CNN performs local feature learning using 2D spatial kernels. CNN has the advantage of local spatial connectivity and shared weights, which helps the convolution layer to perform kernel learning. Batch Normalization is performed after the convolution layer to normalize the activations of each batch by maintaining the mean activation close to zero and standard deviation close to one. The activation function used is Exponential Linear Unit (ELU). Contrary to other activation functions, ELU has negative values too, which pushes the mean of the activations closer to zero, thus helping to speed up the learning process and improving performance \cite{clevert2015fast}. Max pooling is used to make the feature maps robust to noise and distortion and also helps in reducing the number of trainable parameters in the subsequent layers by reducing the size of the feature maps.

Global feature learning is performed using an LSTM layer. The output from the last LFLB is passed on to an LSTM layer to learn the long-term contextual dependencies. Sequences of high-level representation obtained from the CNN+LSTM architecture is passed on to an attention layer whose job is to focus on the emotion salient parts of the feature maps since not all frames contribute equally to the representation of the speech emotion. The attention layer generates an utterance level representation, obtained by the weighted summation of the high-level sequence obtained from CNN+LSTM architecture with attention weights obtained in a trainable fashion \cite{chen20183}. The utterance level attentive representations are passed to a fully-connected layer and then to a softmax layer to map the representations to the different emotion classes.

The number of convolutional kernels in first and second LFLB is 64, and for the third and fourth, LFLB is 128. The size of the convolutional kernels is $ 3 \times 3$  with a stride of $1 \times 1$ for all the LFLBs. The size of the kernel for the max-pooling layer is $ 2 \times 2$ for the first two LFLB and $4 \times 4 $ for the latter two LFLBs. The size of the LSTM cells is 128.

\begin{figure}[!t]
    \subfloat[]{\label{TSNE plot untrained}\includegraphics[height=4cm,width=0.5\textwidth]{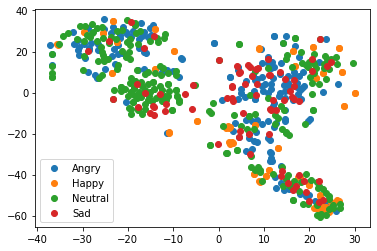}} \\
    \subfloat[]{\label{TSNE after LSTM}\includegraphics[height=4cm,width=0.5\textwidth]{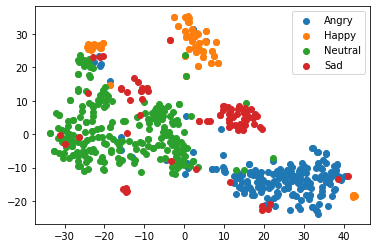}}\\
    \subfloat[]{\label{TSNE after Attention}\includegraphics[height=4cm,width=0.5\textwidth]{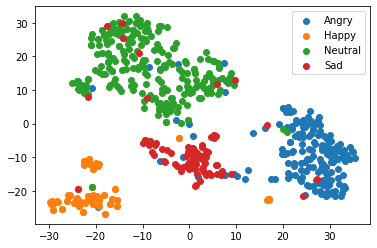}}
    \caption{Distribution of training utterances of IITG Politician Speech Dataset using TSNE with mel-spectrogram as input. \ref{TSNE plot untrained} displays the untrained distribution, \ref{TSNE after LSTM} displays the distribution after the LSTM layer and \ref{TSNE after Attention} displays the distribution after Attention layer.}
    \label{TSNE plot for Politician speech}
\end{figure}

\section{Experimental Evaluation}
\begin{figure*}[!t]
  \centering
  \includegraphics[ height =6cm , width=0.7\linewidth ]{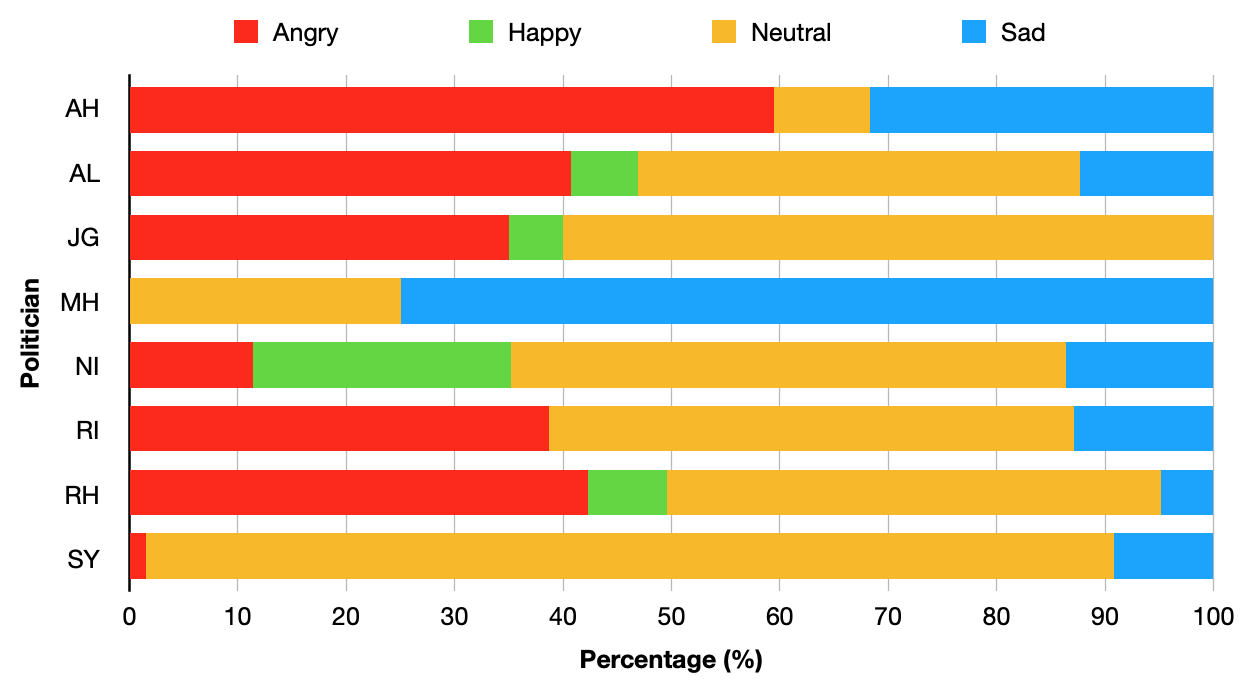}
  \caption{Percentage of different emotions present in a speech of Indian Politicians of the IITG Political Speech Corpus}
  \label{fig:politician_percent}   
\end{figure*}

The experiments are performed in two stages. The first stage of the experiment is to generate class probabilities by performing Speech Emotion Recognition (SER) using the architecture discussed in the above section. The second stage of the experimental evaluation is concerned with the analysis of the duration of different emotions in a politician's speech.

For input to the Attentive CNN+LSTM architecture, the mel-spectrogram is computed from the raw speech files. Since the input to CNN requires equal-sized files, the speech files are either zero-padded or truncated to the same size. A frame size of 25 ms and a frame shift of 10 ms is used to compute the mel-spectrograms. Training of the model is performed in a five-fold cross-validation scheme, and Unweighted Accuracy(UA) is used as the performance measure. The cross-entropy loss function is used in adherence with Adam optimizer to train the model.

The model achieves a recognition performance of UAR $78.95\%$ with the proposed architecture. Figure \ref{fig:confusion_mat} presents the confusion matrix for the predicted emotions of the four emotion classes of the IITG Political speech dataset. Due to less number of samples in the sad emotion category, the recognition performance is degraded for that category. Happy emotion category is clearly recognized by the model. Figure \ref{TSNE plot for Politician speech} presents the distribution of the untrained data, trained data after the LSTM layer, and trained data after the attention layer. The plots show the capability of the architecture in clustering emotions with considerable improvement with the addition of the attention layer.

From data and statistics of various National and State elections in India\cite{elecomm2020, elecomm2019, elecomm2004, elecomm1999}, we made the following observations. 

NI had a vote share of 63.60\%, with a victory margin of 45.20\%. This impressive result can be attributed to his emotionally expressive speeches. The emotional content of his speech is significant in all four emotional categories. 
RH had a vote share of 56.64\%, with a victory margin of 31.07\%. This success is backed by his significant content of anger emotion and other content of few emotions. 
AH had a vote share of 69.58\%, with a victory margin of 43.32\%. The massive vote share and margin are credited to his hugely anger dominated speeches, which compensates for his lack of happy emotion.
JG had a vote share of 33.70\%, with a victory margin of 8.58\%. Although substantial anger can be seen in his speeches, the absence of sad emotional content and poor happiness content could be the reason for his poor vote share and margin. 
RI had a vote share of 64.64\%, with a victory margin of 39.51\%. The electoral success of RI is supported by an alloy of angry and sad emotional content in his speeches. However, a lesser quantity of anger emotion and a dearth of happy emotion elucidate the weak margin.
AL had a vote share of 61.10\%, with a victory margin of 28.35\%. The high vote share can be a result of well balanced emotional content. 
SY had a vote share of a meager 1.62\%, with a huge defeat margin of 54.39\%. An extreme lack of emotion in his oration accounts for this result.
MH, with a vote share of 46.25\%, lost with a defeat margin of 6.00\%. Although a considerable measure of sad emotion is seen in his speeches, the small vote share can be attributed to the lack of potent emotional factors like anger or happiness. 

In summary, emotional content in speeches plays a pivotal role in mobilizing support and success in elections. Amongst the four emotional categories, anger was found to be the most potent political emotion. This is because, in a political scenario, the display of anger in oration make individuals more optimistic about the future, particularly when voters are frustrated by the absence of desired economic and social outcomes like rapid economic growth, reduction in poverty, unemployment, corruption, violence, terrorism, etc. Similarly, happiness is also a vital emotion in an electoral context, as it not only curbs fear \& anxiety but also offers hope for an improved future and adds to a political candidate’s appeal. In the Indian context, sad emotion in speeches may manifest through fear and anxiety amongst the masses. Although sad emotion has a role, it is of lesser importance than other emotions. Furthermore, the most effective speeches in a political context are those that have a mix of all emotions - anger, happiness, and sadness in order of importance.

Our findings are in consonance with the earlier studies concerning the electoral campaigns in the US and point towards the universal application of emotions and their success in politics. SER can, therefore, provide us a metric to gauge the effectiveness of a politician’s speech and the likelihood of success and failure.

\section{Conclusions}

In this work, we presented a brief analysis of the different emotions present in the speeches of politicians in the Indian context. A deep learning framework, CNN+LSTM with attention mechanism, is used to model the emotions using the IITG Politician Speech Dataset. The Dataset is collected from publicly available speeches of Indian politicians on youtube. The speech utterances are divided and annotated by 4 annotators, and a confidence score is provided for each emotion category based on the perceptual evaluation. The performance of the model in predicting the emotions from speech utterances outperforms the perceptual evaluation. An assessment of the percentage of different emotion classes in each of the politician's speech is presented with a brief analysis of the results in accordance with previous results presented in the literature.

\bibliographystyle{IEEEtran}

\bibliography{mybib}

\begin{thebibliography}{10}
\providecommand{\url}[1]{#1}
\csname url@samestyle\endcsname
\providecommand{\newblock}{\relax}
\providecommand{\bibinfo}[2]{#2}
\providecommand{\BIBentrySTDinterwordspacing}{\spaceskip=0pt\relax}
\providecommand{\BIBentryALTinterwordstretchfactor}{4}
\providecommand{\BIBentryALTinterwordspacing}{\spaceskip=\fontdimen2\font plus
\BIBentryALTinterwordstretchfactor\fontdimen3\font minus
  \fontdimen4\font\relax}
\providecommand{\BIBforeignlanguage}[2]{{%
\expandafter\ifx\csname l@#1\endcsname\relax
\typeout{** WARNING: IEEEtran.bst: No hyphenation pattern has been}%
\typeout{** loaded for the language `#1'. Using the pattern for}%
\typeout{** the default language instead.}%
\else
\language=\csname l@#1\endcsname
\fi
#2}}
\providecommand{\BIBdecl}{\relax}
\BIBdecl

\bibitem{riker1968theory}
W.~H. Riker and P.~C. Ordeshook, ``A theory of the calculus of voting,''
  \emph{American political science review}, vol.~62, no.~1, pp. 25--42, 1968.

\bibitem{parker2010vote}
M.~T. Parker and L.~M. Isbell, ``How i vote depends on how i feel: The
  differential impact of anger and fear on political information processing,''
  \emph{Psychological Science}, vol.~21, no.~4, pp. 548--550, 2010.

\bibitem{weber2013emotions}
C.~Weber, ``Emotions, campaigns, and political participation,'' \emph{Political
  Research Quarterly}, vol.~66, no.~2, pp. 414--428, 2013.

\bibitem{valentino2011election}
N.~A. Valentino, T.~Brader, E.~W. Groenendyk, K.~Gregorowicz, and V.~L.
  Hutchings, ``Election night’s alright for fighting: The role of emotions in
  political participation,'' \emph{The Journal of Politics}, vol.~73, no.~1,
  pp. 156--170, 2011.

\bibitem{searles2017use}
K.~Searles and T.~Ridout, ``The use and consequences of emotions in politics,''
  \emph{Emotion Researcher, ISRE’s Sourcebook for Research on Emotion and
  Affect}, 2017.

\bibitem{yang2018predicting}
Z.~Yang and J.~Hirschberg, ``Predicting arousal and valence from waveforms and
  spectrograms using deep neural networks.'' in \emph{Interspeech}, 2018, pp.
  3092--3096.

\bibitem{li2018attention}
P.~Li, Y.~Song, I.~V. McLoughlin, W.~Guo, and L.-R. Dai, ``An attention pooling
  based representation learning method for speech emotion recognition,'' 2018.

\bibitem{busso2008iemocap}
C.~Busso, M.~Bulut, C.-C. Lee, A.~Kazemzadeh, E.~Mower, S.~Kim, J.~N. Chang,
  S.~Lee, and S.~S. Narayanan, ``Iemocap: Interactive emotional dyadic motion
  capture database,'' \emph{Language resources and evaluation}, vol.~42, no.~4,
  p. 335, 2008.

\bibitem{wagner2018deep}
J.~Wagner, D.~Schiller, A.~Seiderer, and E.~Andr{\'e}, ``Deep learning in
  paralinguistic recognition tasks: Are hand-crafted features still relevant?''
  in \emph{Interspeech}, 2018, pp. 147--151.

\bibitem{pandey2019emotion}
S.~K. Pandey, H.~Shekhawat, and S.~Prasanna, ``Emotion recognition from raw
  speech using wavenet,'' in \emph{TENCON 2019-2019 IEEE Region 10 Conference
  (TENCON)}.\hskip 1em plus 0.5em minus 0.4em\relax IEEE, 2019, pp. 1292--1297.

\bibitem{sarma2018emotion}
M.~Sarma, P.~Ghahremani, D.~Povey, N.~K. Goel, K.~K. Sarma, and N.~Dehak,
  ``Emotion identification from raw speech signals using dnns.'' in
  \emph{Interspeech}, 2018, pp. 3097--3101.

\bibitem{chen20183}
M.~Chen, X.~He, J.~Yang, and H.~Zhang, ``3-d convolutional recurrent neural
  networks with attention model for speech emotion recognition,'' \emph{IEEE
  Signal Processing Letters}, vol.~25, no.~10, pp. 1440--1444, 2018.

\bibitem{zhao2019speech}
J.~Zhao, X.~Mao, and L.~Chen, ``Speech emotion recognition using deep 1d \& 2d
  cnn lstm networks,'' \emph{Biomedical Signal Processing and Control},
  vol.~47, pp. 312--323, 2019.

\bibitem{clevert2015fast}
D.-A. Clevert, T.~Unterthiner, and S.~Hochreiter, ``Fast and accurate deep
  network learning by exponential linear units (elus),'' \emph{arXiv preprint
  arXiv:1511.07289}, 2015.

\bibitem{elecomm2020}
E.~C. of~India, ``{General Election to Vidhan Sabha trends and result Feb 2020
  - NCT of Delhi - New Delhi constituency},''
  \url{http://results.eci.gov.in/DELHITRENDS2020/ConstituencywiseU0540.htm?ac=40},
  2020, [Online; accessed 14-March-2020].

\bibitem{elecomm2019}
------, ``{Constituency wise detailed result of General Election 2019},''
  \url{https://eci.gov.in/files/file/10929-33constituency-wise-detailed-result/},
  2019, [Online; accessed 14-March-2020].

\bibitem{elecomm2004}
------, ``{General Election Archive(1951 - 2004), General Election 2004(Vol.
  III)},''
  \url{https://eci.gov.in/files/file/4126-general-election-2004-vol-i-ii-iii/},
  2018, [Online; accessed 14-March-2020].

\bibitem{elecomm1999}
------, ``{General Election Archive(1951 - 2004), General Election 1999(Vol.
  III)},''
  \url{https://eci.gov.in/files/file/4125-general-election-1999-vol-i-ii-iii/},
  2018, [Online; accessed 14-March-2020].

\end{thebibliography}


\end{document}